\documentclass{article}
\usepackage[flushleft]{threeparttable}
\usepackage{xr-hyper}
\usepackage{hyperref}
\usepackage[detect-all]{siunitx}
\usepackage{import,gensymb}
\usepackage{colortbl}
\usepackage{booktabs}
\usepackage{array,multirow}

\usepackage{amsmath}
\usepackage{amsthm}
\usepackage{amssymb}
\usepackage{amsfonts}
\usepackage{dsfont}
\usepackage{accents}
\usepackage{mathtools}
\usepackage{algorithm}
\usepackage{algorithmic}
\usepackage{graphicx}
\usepackage{subfigure}
\usepackage{float}
\usepackage{caption}
\usepackage{tikz,pgfplots}
\begin{document}
\title{Quantitative reconstruction of defects in multi-layered bonded composites using fully convolutional network-based ultrasonic inversion}
\author{Jing Rao,
        Fangshu Yang,
         Huadong Mo,\\ 
         Stefan Kollmannsberger,
				 Ernst Rank
\thanks{The work was supported by the start-up grant from UNSW Canberra, and the German Research Foundation (DFG) under the Grant No. KO4570/1-1 and Ra624/29-1.~(Corresponding author: Jing Rao)
Jing Rao is with School of Engineering and Information Technology, University of New South Wales, Canberra, ACT 2600, Australia.(e-mail: \url{jing.rao@adfa.edu.au}
Fangshu Yang is with the School of Mathematics, Institute of Artificial Intelligence, Harbin Institute of Technology, Harbin 150001, China. 
Huadong Mo is with School of Engineering and Information Technology, University of New South Wales, Canberra, ACT 2600, Australia.
Stefan Kollmannsberger is with the Chair of Computational Modelling and Simulation, Technical University of Munich, Arcisstr. 21, 80333 Munich, Germany.
Ernst Rank is with Chair of Computational Modelling and Simulation, Technical University of Munich, Arcisstr. 21, 80333 Munich, Germany, and also with Institute for Advanced Study, Technical University of Munich, Lichtenbergstr. 2a, 85748 Garching, Germany.}}



\maketitle
\begin{abstract}
Ultrasonic methods have great potential applications to detect and characterize defects in multi-layered bonded composites. However, it remains challenging to quantitatively reconstruct defects, such as disbonds and kissing bonds, that influence the integrity of adhesive bonds and seriously reduce the strength of assemblies. In this work, an ultrasonic method based on the supervised fully convolutional network (FCN) is proposed to quantitatively reconstruct defects hidden in multi-layered bonded composites. In the training process of this method, an FCN establishes a non-linear mapping from measured ultrasonic data to the corresponding velocity models of multi-layered bonded composites. In the predicting process, the trained network obtained from the training process is used to directly reconstruct the velocity models from the new measured ultrasonic data of adhesively bonded composites. The presented FCN-based inversion method can automatically extract useful features in multi-layered composites.~Although this method is computationally expensive in the training process, the prediction itself in the online phase takes only seconds. The numerical results show that the FCN-based ultrasonic inversion method is capable to accurately reconstruct ultrasonic velocity models of the high contrast defects, which has great potential for online detection of adhesively bonded composites.
\end{abstract}


\section{Introduction}
Adhesively bonded multi-layer structures are of interest in aerospace, marine, train, nuclear, and offshore industries~\cite{Quaegebeur2012,Pyzik2021,Giri2019,Higgins2000,Chester1999}. The advantages of adhesive bonds are that they are quite easy to manufacture and to assemble at relatively low cost - and that they are able to distribute mechanical loads evenly and, thus, decrease stress concentration~\cite{Cerniglia2008}. The quality of adhesive bonds can be compromised by external loading, various types of harsh environment conditions, or natural aging, leading to defects (e.g., disbonds and kissing bonds) in the adhesive layer or decrease of the adhesive strength~\cite{Marques2008,Yan2012}. Therefore, the accurate and efficient assessment of the bonding quality is of prime importance to structural integrity and reliability. Non-destructive evaluation (NDE) methods have been shown to be useful for evaluating the health conditions of the adhesively bonded structures, such as thermography~\cite{Zhao2016}, shearography~\cite{Taillade2011} and ultrasound~\cite{Zhang2019,Hung2000}.

Thermography and shearography are full-field, non-contact NDE methods that show potential as practical tools to determine the adhesive bond quality ~\cite{Hung2007}. They are often used to address disbonds/delaminations in composite structures~\cite{Shang1991}. However, they are based on different defect detection mechanisms. Thermography uses energy to heat the surface, and it detects infrared radiation from the surface to measure the surface temperature response to thermal excitation. The fundamental mechanism for the detection of subsurface defects is thermal diffusion of energy from the surface to the interior, which is why the limitations imposed by diffusion affect the minimum detectable defect size. Shearography measures the specimen's mechanical response to stress loading such as thermal, vacuum and mechanical excitations~\cite{Hung2007}. It should avoid introducing excessive rigid-body motion between specimen and measurement system, which causes speckle decorrelation and deteriorates fringe visibility, thus reducing the detection sensitivity of defects. Due to these limitations, these two techniques are generally inappropriate for detecting deep defects hidden in adhesively bonded structures.

The detection of deeper disbonds and voids can be provided by ultrasonic bulk wave testing, which can be broadly divided into techniques applied in the time domain and in the frequency domain~\cite{Allin2003}. For example, through-transmission and analysis of pulse-echo signals in the time domain~\cite{Freemantle1997,Goglio1999}, as well as the fundamental through thickness resonance frequency~\cite{Allin2003} have been used for detecting disbonds, albeit at a degree of quantitative assessment of the multi-layer structures. Nevertheless, when extended to the quantitative detection of disbond and kissing bond defects, little work has been carried out to achieve quantitative reconstructions of these high contrast defects in adhesively bonded structures by building accurate ultrasonic velocity models.

There are two major groups of velocity model building techniques: making use of the focusing properties of migration~\cite{Al-Yahya1989,Deregowski1990}, and using some forms of information such as traveltime extracted from the data~\cite{Sattleger1981,Williamson1990,Tarantola1986}. Many techniques in these two major groups may require either repeated application of the migration or time-consuming picking of traveltime information in the measured data~\cite{Kluver2007}. Therefore there is a need for an approach that can exceed the accuracy of these conventional methods by using all information contained in the measured data while at the same time avoiding the computational complexity and inversion constraints.

Deep learning, a subset area of machine learning, shows potential to alleviate these restrictions. Deep learning leverages all signal content in the data for predicting models, helps to avoid incorrect assumptions in the physics of wave propagation, and offers computational advantages over conventional inversion methods~\cite{Kim2018}. In deep learning, a convolutional neural network (CNN), which is one of the most popular choices of deep neural networks~\cite{LeCun2010}, is able to approximate the non-linear mapping from the input space to the output space. For a detailed survey on CNNs, we refer to~\cite{Krizhevsky2012,Rawat2017,Simard2003,Voulodimos2018}. Here, the summary only includes terms, principles, and basic properties that are necessary to understand the specific topic of this paper. CNNs have demonstrated remarkable ability for images and label maps~\cite{Zheng2015}, images and text~\cite{Ghosh2019}, as well as different types of images~\cite{Zhou2016,Godard2017}, especially for inverse problems, such as model/image reconstruction~\cite{Jin2017,yang2021robust} and image super-resolution~\cite{Dong2015}. This state-of-the-art development opens up new perspectives for signal inversion and velocity model building, and several works have already made progress in this regard. For example, Araya-Polo $et \ al.$~\cite{Araya-Polo2018} used CNNs to reconstruct velocity models directly from seismic data. Li $et \ al.$~\cite{Li2020} further proposed an end-to-end CNN to consider uncertain relationships between seismic data and velocity models, and the time-varying property of seismic data. In general, the training process of CNNs is computationally expensive, but it only needs to be performed once up front. After training, the computational costs of predicting the velocity reconstruction are negligible, rendering the overall computational costs to be only a fraction of conventional inversion methods~\cite{Kim2018}.

When a standard multi-layer perceptron is used in the CNN, meaning fully-connected layers, CNNs become expensive to compute due to the large number of dimensions involved, and too many parameters in fully-connected layers slow down the training speed of the network~\cite{Jiang2019}. Besides, traditional CNNs are not very suitable to identify highly complex settings containing different backgrounds and a lot of overlap. To address these issues, a fully convolutional network (FCN) was proposed~\cite{Long2015}, replacing the fully-connected layers with only convolutional layers, and it can better preserve the neighbourhood information in the pixel-wise outputs~\cite{Nie2016}. Furthermore, it has been observed that a modified FCN which has an encoder-decoder structure yields more precise predictions~\cite{yang2020deep}. It consists of a contracting path which is used to capture the useful features and a symmetric expanding path that enables precise localization or reconstruction, showing good performance in velocity model building.

\begin{figure}[!htbp]
	\center
		\includegraphics[width=\columnwidth]{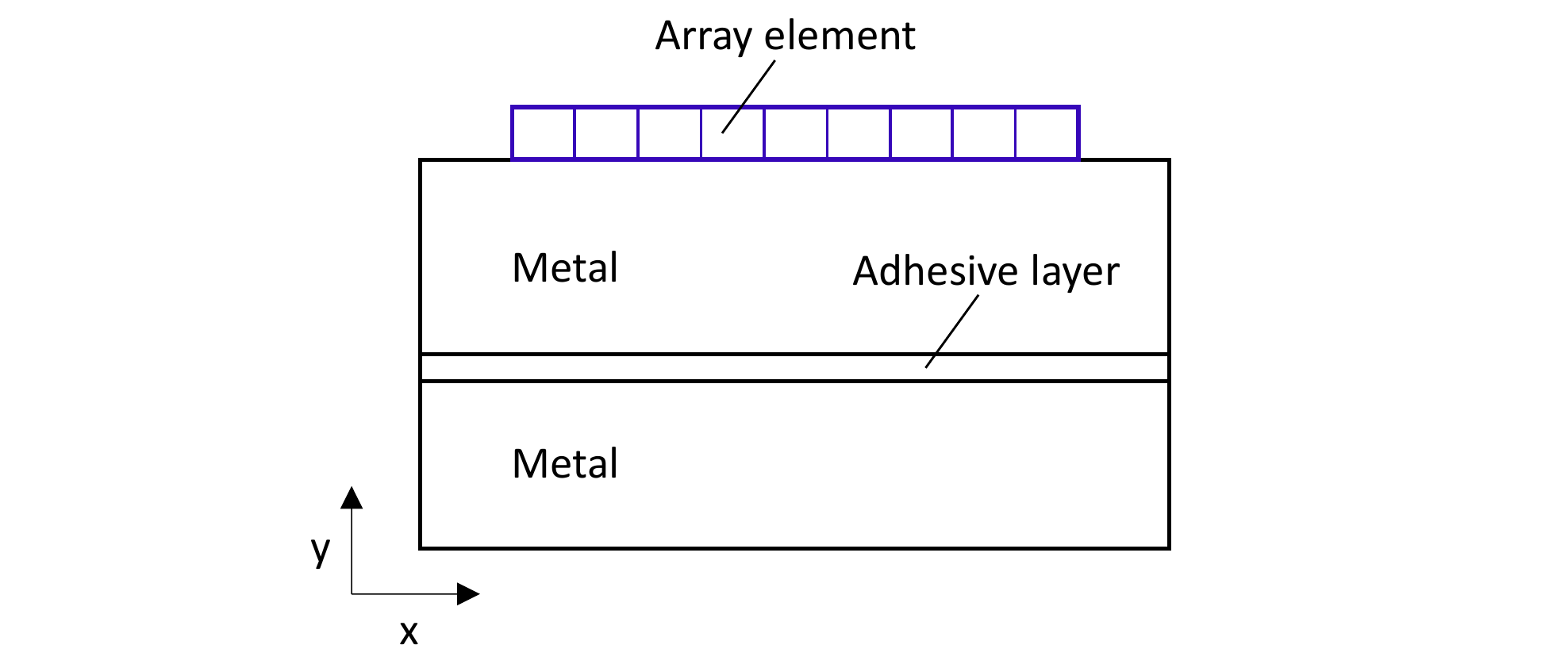}
		\caption{Schematic diagram of an adhesively bonded structure using an ultrasonic linear phased array to collect the FMC data. \label{fig:general_configure}}
\end{figure}

In this work, we propose a new FCN-based encoder-decoder network to directly reconstruct the ultrasonic longitudinal wave (L-wave) velocity models of multi-layered bonded composites containing high contrast defects from raw ultrasonic data, instead of performing the inversion of ultrasonic full waveforms~\cite{Rao2016}. In the training process, the full matrix capture (FMC) is used as input to the neural network. There, each element transmits a signal in turn, which is then measured by all of the elements in an ultrasonic linear phased array, as shown in Fig.~\ref{fig:general_configure}. In other words, FMC contains a set of ultrasonic signals for all transmitter-receiver combinations in the linear array, giving a complete data set. The proposed FCN-based network is used to approximate the unknown non-linear mapping from the FMC data to the corresponding ultrasonic L-wave velocity model. In the predicting process, the trained network can be utilized to predict the unknown adhesively bonded structures using new FMC data. Compared with conventional inversion methods, no initial velocity models and less human intervention are involved in the proposed FCN-based inversion method.

The remainder of this paper is organized as follows. First, we introduce the formulation of the forward and inversion problems and briefly review FCN in Section II. In Section III, the FCN-based ultrasonic inversion method is presented in detail. Then, in Section IV, we will provide numerical examples of adhesively bonded structures with various defects, including data preparation and inversion results, followed by a discussion of the effectiveness of this method. A general discussion follows in Section V, and the final conclusions are summarized in Section VI.

\section{Background theory}
\subsection{Problem formulation}\label{sec:Inversion problem}

Due to limitations of the computational resources, we restrict our description in this paper to a two-dimensional (2D) setting without losing generality. In the 2D setting, it is assumed that the adhesively bonded structure is long in the z-direction (see Fig.~\ref{fig:general_configure}) with through-wall defects, and the linear phased array is also long in this direction. Therefore, the time-domain 2D acoustic wave equation in a constant density media can be given by
\begin{equation}\label{eq:forward_modelling}
\begin{aligned}
& \frac{1}{v^2{(x,y)}} \cdot \frac{\partial^2{{d_{cal}(x,y,t)}}}{\partial{t}^2} \\ 
& =\frac{\partial^2{{d_{cal}(x,y,t)}}}{\partial{x}^2}+\frac{\partial^2{{d_{cal}(x,y,t)}}}{\partial{y}^2}+s(x,y,t),
\end{aligned}
\end{equation}
where $v(x,y)$ represents the L-wave velocity at a spatial position $(x,y)$. $d_{cal}(x,y,t)$ represents the pressure (or displacement) wavefield, and the time is denoted by $t$ and the source by $s(x,y,t)$. If a forward problem is considered, $v(x,y)$ is expected to be given, whereas $d_{cal}(x,y,t)$ is the unknown wavefield. 

The relationship between the pressure wavefield and the velocity is non-linear and Eq.~(\ref{eq:forward_modelling}) can be written compactly through the operator $F$, defined as
\begin{equation}\label{eq:forward_matrix_form}
\centering
{d}_{cal}=F({v}).
\end{equation}

\noindent The aim of the classical inversion method is to identify a velocity model ${v}$ by minimizing the objective function 
\begin{equation}\label{eq:objective_function}
\arg \min\limits_{{v}} f({v}) =
\arg \min\limits_{{v}} \frac{1}{2} {\left \|  F({v}) - {d}  \right\|^2_2},
\end{equation}
where ${d}$ represents pressure wavefields at every receiver and every time step which are captured through the FMC (i.e., measured data). $f({v})$ is the data residual between the modelling data computed in the model and the measured data. $\left \| {\cdot} \right\|_2$ denotes the $L_2$ norm. Note that this paper serves to propose a conceptual approach to reconstruct velocity models, and therefore physical experiments are replaced by numerical simulations to obtain the measured data.

First-order descent methods can be used to solve the optimization problem Eq.~(\ref{eq:objective_function}), requiring the computation of the gradient of the objective function, i.e., $\nabla f({v})$. For example, Plessix et al.~\cite{Plessix2006} applied an adjoint-state method to compute the gradient without explicitly building the sensitivity matrix and then used an iterative optimization algorithm to minimize the objective function. Because of the non-linear operator $F$ and the incomplete and inaccurate measured data ${d}$, it is difficult to obtain the unique precise velocity model ${v}$, which is why minimizing the above objective function is generally an ill-posed inverse problem.

\subsection{Conventional fully convolutional network}\label{sec:review_of_FCN}

\begin{figure*}[!htbp]
	\center
		\includegraphics[width=0.8\textwidth]{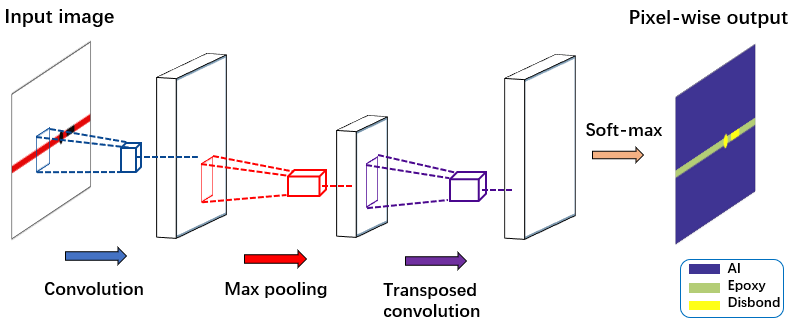}
		\caption{A sketch of a simple FCN architecture, including a convolutional layer, a pooling layer and a transposed convolutional layer. \label{fig:general_FCN}}
\end{figure*}

Based on universal approximation properties of neural networks to continuous functions~\cite{Hornick1989}, fully convolutional networks (FCN) have been successfully applied to solve ill-posed inverse problems. FCNs use convolutional layers to achieve end-to-end learning, instead of fully-connected layers used in CNNs~\cite{Jin2017}. A simple FCN architecture is shown in Fig.~\ref{fig:general_FCN}. Images are used as inputs, followed by a convolutional layer and a pooling layer. After applying the max pooling layer, the sizes of the feature maps are compressed. Afterward, the transposed convolutional layer is employed to bring the output sizes to the original input sizes. Finally, a soft-max function is used to classify the feature maps at the pixel level (pixel-wise output). This FCN architecture can also be interpreted as an encoder-decoder path. More details of the FCN can be found in~\cite{Long2015}. The conventional FCN can be formulated as
\begin{equation}\label{eq:FCN_expression}
\mathbf{Y}=Net(\mathbf{X};\mathbf{\Theta})=S(\mathbf{w}_2*(M(R(\mathbf{w}_1*\mathbf{X}+\mathbf{b}_1)))+\mathbf{b}_2),
\end{equation}
where $Net$ represents the FCN-based network. $\mathbf{X}$ and $\mathbf{Y}$ are the input images and pixel-wise outputs (2D), respectively. $\mathbf{\Theta}$ represents the learnable parameters of the network, including the weights $\mathbf{w}_1$ and $\mathbf{w}_2$, as well as the biases $\mathbf{b}_1$ and $\mathbf{b}_2$. $R$ denotes a non-linear activation function like Tanh, Sigmoid and Rectified Linear Unit (ReLU)~\cite{Dahl2013}. $M$ is the sub-sampling function, e.g., max pooling, and $S$ is the soft-max function. The convolutional operation is denoted by $*$.

\section{FCN-based ultrasonic inversion}

\begin{figure*}[!htbp]
	\center
		\includegraphics[width=0.8\textwidth]{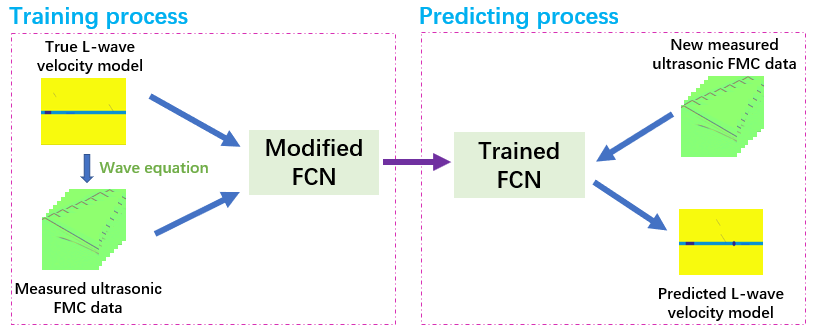}
		\caption{Structure of FCN-based ultrasonic inversion method. \label{fig:ultrasonic_modified_FCN}}
\end{figure*}

In this paper, our goal is to reconstruct a 2D ultrasonic L-wave velocity model (outputs in the model domain) by directly using the measured ultrasonic FMC data (inputs in the data domain) based on the linear phased array, and we therefore propose a FCN-based ultrasonic inversion method. The basic idea of this method is to establish the mapping between inputs and outputs, which can be given by
\begin{equation}\label{eq:modified_expression}
\mathbf{\hat{v}}=Net(\mathbf{d};\mathbf{\Theta}),
\end{equation}
where $\mathbf{\hat{v}}$ represents the 2D L-wave velocity models predicted by the network, and $\mathbf{d}$ denotes the measured ultrasonic FMC data. This method consists of a training process and a predicting process, as shown in Fig.~\ref{fig:ultrasonic_modified_FCN}. Before training, the true ultrasonic velocity models of adhesive bonded composites with different defects are randomly created and then the measured ultrasonic FMC signals are obtained from simulations. A large amount of measured ultrasonic signals and the corresponding "true" velocity models, i.e., $\{\mathbf{d}_n,\mathbf{v}_n\}_{n=1}^N$, are assembled to train the network. It should be mentioned that although defects in adhesive bonded composites cannot be randomly generated in real-world scenarios, it is possible to prefabricate different types of through-wall defects in the physical laboratory experiments and then collect field test examples. 

\begin{figure*}[!htbp]
	\center
		\includegraphics[width=0.8\textwidth]{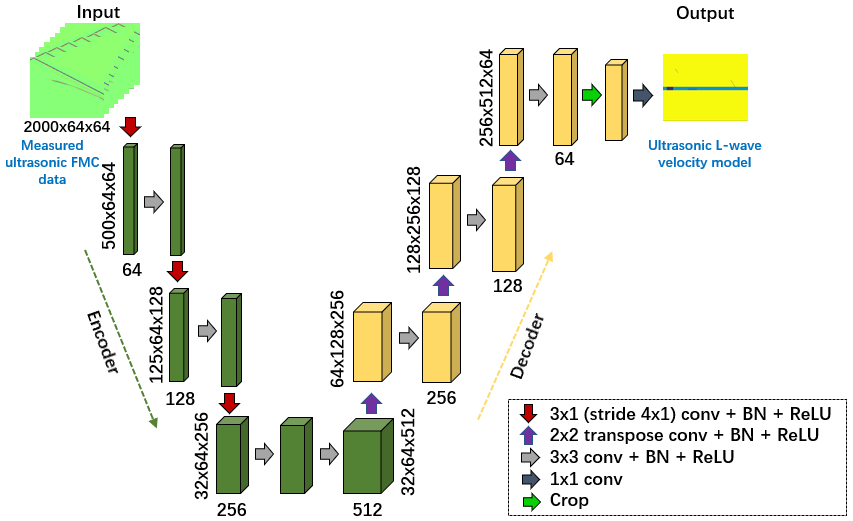}
		\caption{An illustration of the network architecture used in the FCN-based ultrasonic inversion method. Note that conv, BN, and ReLU denote 2D convolution, batch normalization, and Rectified Linear Unit, respectively. \label{fig:Architecture_network}}
\end{figure*}

To achieve the ultrasonic velocity model building directly from the measured ultrasonic FMC data, the conventional FCN architecture (see Fig.~\ref{fig:general_FCN}) is adopted and modified in this work~\cite{badrinarayanan2017segnet}, as shown in Fig.~\ref{fig:Architecture_network}. There are two major modifications in this proposed encoder-decoder architecture of the network to match the phased array ultrasonic testing of adhesively bonded composites. Firstly, the conventional FCN architecture is designed for image segmentation and reads input images in grey or red-green-blue colour channels. In this work, the ultrasonic FMC data, which is described in Section I, is acquired as input. The number of transmitters in each model is utilized as the number of channels for inputs. Secondly, the inputs and the outputs are in the same image domain in the conventional FCN, while the proposed architecture is used to achieve the domain transformation from the data domain to the model domain. 

From Fig.~\ref{fig:Architecture_network}, we can see that the encoder is a down-sampling process that is used to extract feature maps from the input ultrasonic FMC data. Taking the simulated FMC data of $2000 \times 64 \times 64$ (sampling points $ \times $ receivers $ \times $ transmitters) as an example, the time step in this work is 5e-9 s (i.e., the total time is 1e-5 s) and this measured data contains $64 \times 64$ time traces. The feature map obtained through the convolutional operation has 64 channels and its dimensions are $500 \times 64 \times 64$. Then, the number of channels is doubled in each operation of the encoder path. After this, the feature map extracted by the encoder is enlarged by the corresponding decoder (up-sampling process). Finally, a cropping process is added after the last feature map to ensure that the output sizes are the same as the input sizes. 

During the training process, this proposed network is required to fit a non-linear function from the input ultrasonic FMC data to the corresponding velocity models, and thus the network is built by solving the optimization problem
\begin{equation}\label{eq:optimization_FCN}
\mathbf{\hat{\Theta}}=\arg \min\limits_{\mathbf{\Theta}} \frac{1}{mN}  \sum_{n=1}^{N}L(\mathbf{v}_n,Net(\mathbf{d}_n;\mathbf{\Theta})),
\end{equation}
where $m$ denotes the total number of pixels in one velocity model and $N$ is the number of training datasets. The loss function $L$ measures the difference between the true velocity models $\mathbf{v}_n$ and the predicted velocity models $\mathbf{\hat{v}}_n$. In our work, we set $L(\mathbf{v}_n, \mathbf{\hat{v}}_n) = \sum(\beta \times (-\log(\mathbf{s})) \times|\mathbf{v}_n-\mathbf{\hat{v}}_n |$), where $\mathbf{s}$ represents the matrix of pixel-wise probability~\cite{jadon2020survey}, namely the percentage of pixels representing distinct colours (i.e., representing different cracks, disbonds, aluminium layers, etc.) in the total number of pixels. $\beta$ is a user-defined weight matrix based on the matrix $\mathbf{s}$, in order to make a trade-off between different colours. $\rm{log}(\mathbf{s})$ is the logarithm of each value in the matrix $\mathbf{s}$, and it serves to improve the sensitivity of the small values in this matrix. Otherwise, the small percentage of coloured pixels (e.g., representing small defects) would be very difficult to detect. $|\cdot|$ denotes the absolute value. 
Note that $L$ used in Eq.~(\ref{eq:optimization_FCN}) is different from $f$ used in the classical inversion method (Eq.~(\ref{eq:objective_function})), in which the function measures the squared difference between the modelling data and the measured data. Because of the relatively large number of the training datasets $N$, we use a small subset of the whole training dataset, namely the mini-batch size $h$, in each iteration to calculate $L_h$. Then, Eq.~(\ref{eq:optimization_FCN}) can be rewritten in the following form
\begin{equation}\label{eq:optimization_FCN_sub}
\begin{aligned}
\mathbf{\hat{\Theta}} = \arg \min\limits_{\mathbf{\Theta}}  \frac{1}{mh} \sum_{n=1}^{h}L_{h}(\mathbf{v}_n,Net(\mathbf{d}_n;\mathbf{\Theta})). 
\end{aligned}
\end{equation}

We sequentially process small batches of the shuffled training data to ensure one epoch $e$ (i.e., single pass), which requires exactly one forward and one backward pass through all of the training data~\cite{Justus2018}. In this work, the Adam algorithm~\cite{Kingma2015} is used to update learnable parameters to minimize the objective function as follows:
\begin{equation}\label{eq:learning_parameters_Adam}
{\mathbf{\Theta}}_{(e+1)} = {\mathbf{\Theta}}_{(e)} - \alpha g(\frac{1}{mh} \nabla_{\mathbf{\Theta}} L_h ({\mathbf{d}_n};{\mathbf{\Theta}};{\mathbf{v}_n})),
\end{equation}
where $\alpha$ is a positive learning rate and $g(\cdot)$ represents a function. The gradient of ${L_h}$ is computed using the chain rule to find the derivative of the weights and biases of ${L_h}$. The advantages of the Adam algorithm are easy implementation and computational efficiency. 

In the predicting process, the unknown velocity models can be reconstructed from the new measured FMC data based on the trained network.

The full algorithm of the FCN-based ultrasonic inversion method is outlined in Algorithm \ref{alg1}, which is implemented in PyTorch~\cite{Paszke2017}, and the list of symbols and notations used in Algorithm \ref{alg1} is shown in Table \ref{tab:list_of_symbols}. All computations presented in the following are carried out on a desktop workstation (GeForce GPU, Ubuntu operating system).
\begin{algorithm*}
\caption{\small FCN-based ultrasonic inversion method}
\textbf{Input:} $\{\mathbf{d}_{n}\}_{n=1}^{N}$: measured ultrasonic FMC data, $\{\mathbf{v}_{n}\}_{n=1}^{N}$: ultrasonic L-wave velocity models, $E$: epoch, $\alpha$: learning rate of the Adam algorithm, $h$: batch size, $N$: number of training dataset, $l$:~half of all neural network layers. \\
\textbf{Initialize:} $\mathbf{y}_0=\mathbf{d}$\\
\textbf{Training process}
\begin{enumerate}
 \item Create different ultrasonic velocity models of adhesively bonded composites.
 \item Acquire the measured FMC data from simulations.
 \item Import the measured FMC data into the network and use the Adam to update the learnable parameters as follows:
    \begin{algorithmic}
    \FOR {e=1:E}
    \FOR {j=1:N/h}
    \FOR {i=1:l-1}
    \STATE $\mathbf{y}_{i}\leftarrow \rm{ReLU}(\rm{BN}(\mathbf{w}_{(2i-1)}\ast_{\downarrow} \mathbf{y}_{i-1}+\mathbf{b}_{(2i-1)}))$\\
    \STATE $\mathbf{y}_{i}\leftarrow \rm{ReLU}(\rm{BN}(\mathbf{w}_{(2i)}\ast \mathbf{y}_{i}+\mathbf{b}_{(2i)})) \quad \quad \quad \quad \quad \ \ \ \triangleright{\ \rm{Fig.~({\ref{fig:Architecture_network}})} \ Encoder}$\\
    \ENDFOR
    \STATE $\mathbf{y}_{l}\leftarrow \rm{ReLU}(\rm{BN}(\mathbf{w}_{(2l-1)}\ast \mathbf{y}_{l-1}+\mathbf{b}_{(2l-1)}))$\\
    \FOR {i=l-1:1}
    \STATE $\mathbf{y}_{i}\leftarrow \mathbf{w}_{(2l+3(l-i)-1)}\ast_{\uparrow} \mathbf{y}_{i+1}+\mathbf{b}_{(2l+3(l-i)-1)}  $\\
    \STATE $\mathbf{y}_{i}\leftarrow \rm{ReLU}(\rm{BN}(\mathbf{w}_{(2l+3(l-i))}\ast \mathbf{y}_{i}+\mathbf{b}_{(2l+3(l-i))})) \quad \triangleright{\ \rm{Fig.~({\ref{fig:Architecture_network}})} \ Decoder}$
    \ENDFOR
    \STATE $\mathbf{\hat{v}}\leftarrow \mathbf{w}_{(5l-2)}\ast_{1} \rm{C}(\mathbf{y}_{1})+\mathbf{b}_{(5l-2)}$\\
    \STATE $\rm{Loss}$= $\frac{1}{mh} \sum_{n=1}^{h} L_{h}(\mathbf{v}_n,\mathbf{\hat{v}}_n)$ \quad \quad \quad \quad \quad \quad \quad \quad \quad  $ \ \ \ \ \triangleright{\ \rm{Eq.~({\ref{eq:optimization_FCN_sub}})}}$ \\
    \STATE $\mathbf{\Theta}_{e_{j+1}} \leftarrow \rm{Adam} (\mathbf{\Theta}_{e_{j}},$ \textit{$\alpha$} $,\rm{Loss})$
    \ENDFOR
    \ENDFOR
   \end{algorithmic}
\end{enumerate}
\textbf{Predicting process}
\begin{enumerate}
  \item Import the new measured FMC data into the trained network for the prediction.
\end{enumerate}
\textbf{Output:} predicted ultrasonic velocity model $\mathbf{\hat{v}}$.
\label{alg1}
\end{algorithm*}
\begin{table}[!t]
	\caption{List of symbols and notations used in Algorithm \ref{alg1}}
	\label{tab:list_of_symbols}
	\centering
	{\begin{tabular}{ll}
			\hline
			\hline
		     $\mathbf{\Theta}=\{\mathbf{w},\mathbf{b}\}$ & learnable parameters \\
			\hline
			 $\ast$ & 2D convolution (kernel size $3\times 3$, stride $1\times 1$) \\
			\hline
		    $\ast_{\uparrow}$ & 2D transposed convolution (kernel size $2\times 2$, stride $1\times 1$)  \\
		    \hline
			$\ast_{\downarrow}$ & 2D convolution (kernel size $3\times 1$, stride $4\times 1$)  \\
			\hline
			$\ast_{1}$ & 2D convolution (kernel size $1\times 1$, stride $1\times 1$) \\ 
			\hline
			ReLU & Rectified Linear Unit \\
		    \hline
			Adam & algorithm for stochastic optimization \\
			\hline
			BN & batch normalization \\ 
			\hline
			C & cropping \\ 
			\hline
			\hline
	\end{tabular}}
\end{table}
\section{Numerical models and results}\label{sec:Testing approach}

This section serves to present the data preparation, including ultrasonic velocity models of adhesively bonded composites and modelling procedures for both training and testing datasets. 

\subsection{Data preparation}\label{sec:forward modelling}
\subsubsection{Ultrasonic velocity models for training process}\label{sec:velocity models}

\begin{figure*}[!htbp]
	\center
		\includegraphics[width=0.9\textwidth]{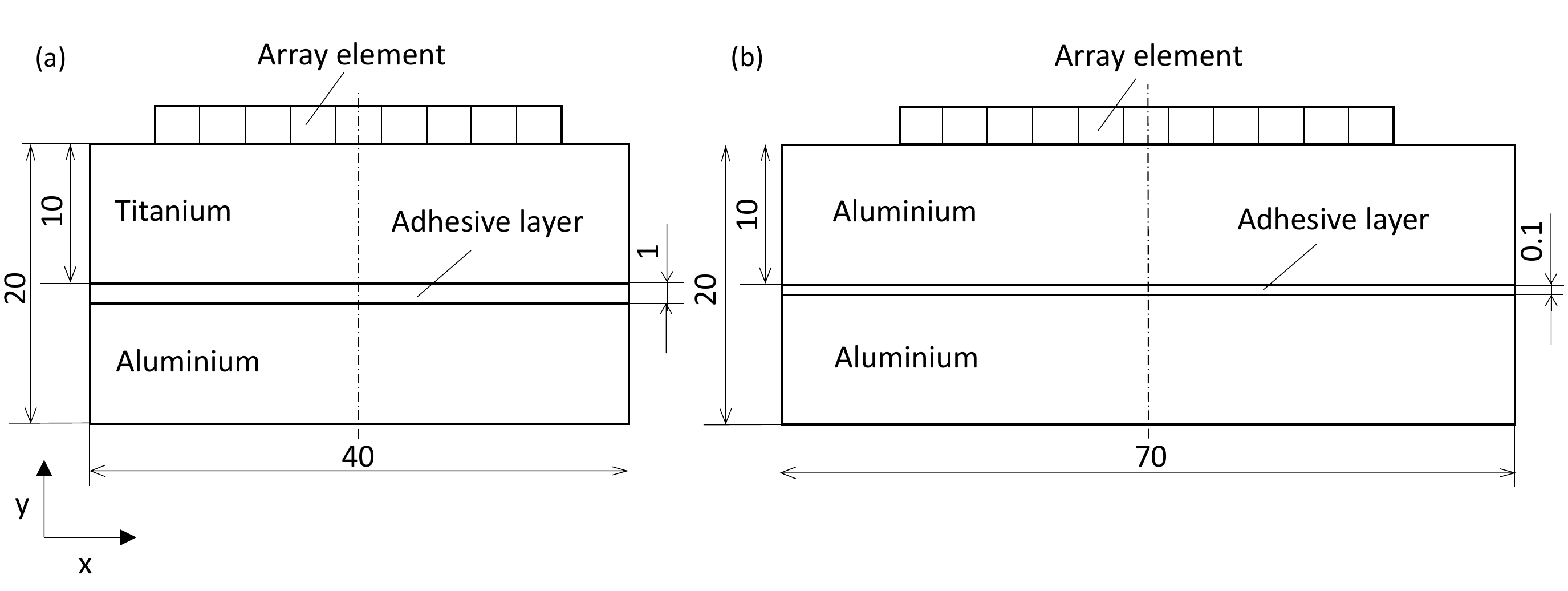}
		\caption{Configurations of (a) Type I with 1 mm-thick adhesive layer and (b) Type II with 0.1 mm-thick adhesive layer. The sizes of Type I and Type II are 40 mm $\times$ 20 mm and 70 mm $\times$ 20 mm, respectively. \label{fig:configuration}}
\end{figure*}

\begin{figure*}[!htbp]
	\center
		\includegraphics[width=0.8\textwidth]{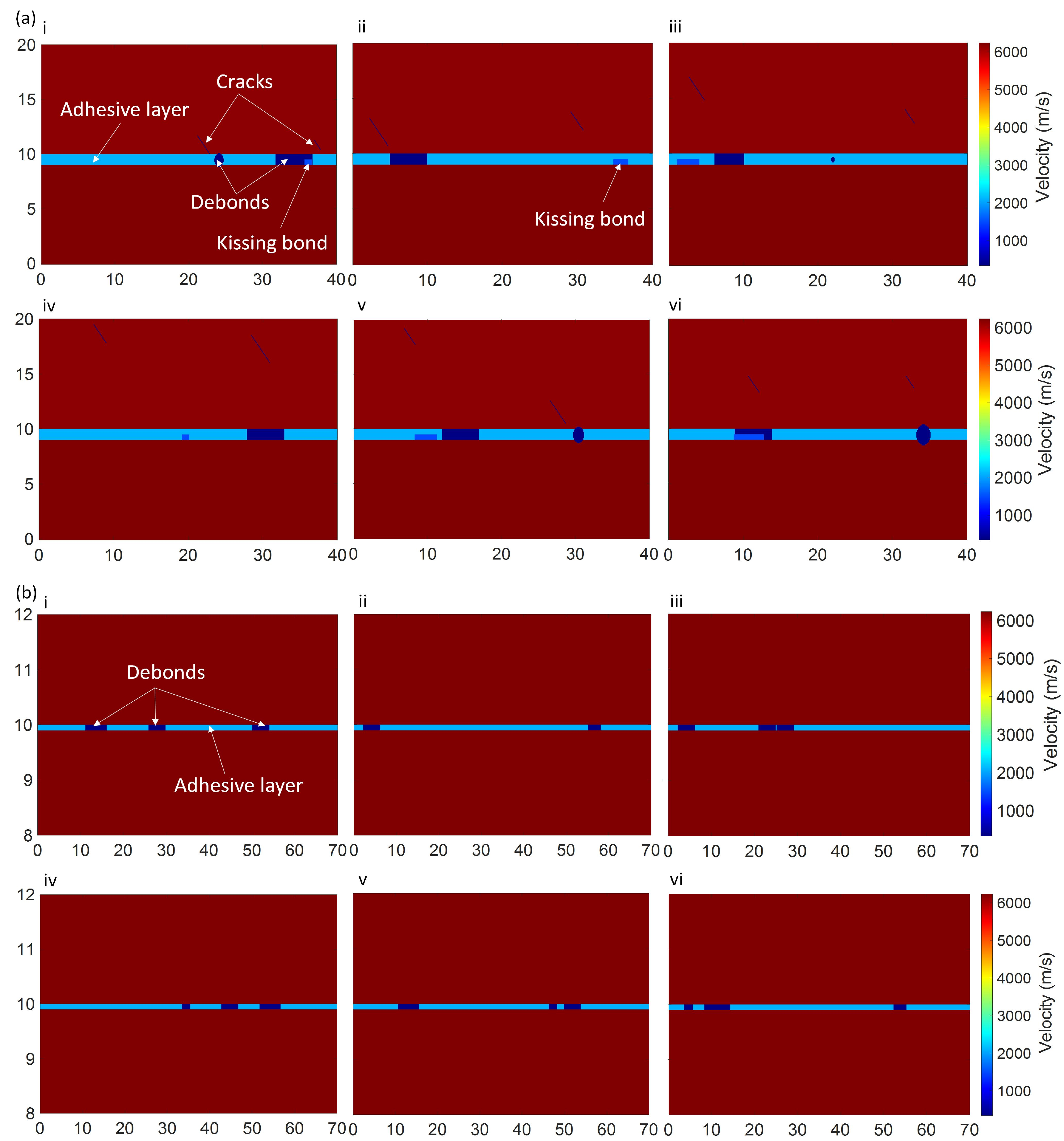}
		\caption{Six representative ultrasonic L-wave velocity models of (a) Type I with debonds, kissing bond and cracks, as well as (b) Type II with multiple debonds. Note that (b) only shows a range from 8 mm to 12 mm in the y-direction. The unit in this figure is mm. \label{fig:random_velocity_models}}
\end{figure*}

To train an efficient network for quantitatively detecting defects in multi-layered bonded composites, a relatively large number of ultrasonic velocity models are randomly generated first. In this work, we consider the adhesively bonded models by bonding two metal layers with an epoxy resin adhesive layer. The typical configurations of Type I and Type II are shown in Figs.~\ref{fig:configuration} (a) and (b), respectively. In the examples of Type I, the dimensions of the top titanium layer are 40 mm $\times$ 10 mm, the bottom aluminium layer is 40 mm $\times$ 9 mm, and the adhesive layer is 40 mm $\times$ 1 mm (see Fig.~\ref{fig:configuration} (a)). Two cracks with the width of 0.1 mm are randomly created in the top layer of titanium. Different sizes and locations of disbond and kissing bond defects in the adhesive layer are also introduced. In Type I, the simulated training dataset contains 800 ultrasonic L-wave velocity models, and 6 typical velocity models are shown in Fig.~\ref{fig:random_velocity_models}(a). In the examples of Type II, the dimensions of the top aluminium layer, the bottom aluminium layer and the adhesive layer are 70 mm $\times$ 10 mm, 70 mm $\times$ 9.9 mm and 70 mm $\times$ 0.1 mm, respectively (see Fig.~\ref{fig:configuration} (b)). Multiple debonds with different lengths are randomly created in the 0.1 mm-thick adhesive layer. In Type II, 800 ultrasonic L-wave velocity models are included in the training dataset, and 6 representative velocity models are shown in Fig.~\ref{fig:random_velocity_models}(b). The ultrasonic L-wave velocities of different materials and defects are shown in Table \ref{tab:velocity_in_different_areas}. 
\begin{table}[!t]
	\caption{Ultrasonic L-wave velocity of different materials and defects}
	\label{tab:velocity_in_different_areas}
	\centering
	{\begin{tabular}{lc}
			\hline
			\hline
			 & Ultrasonic L-wave velocity (m/s) \\
			\hline
			Aluminium  & 6235 \\
			\hline
			Titanium  & 6144 \\
			\hline
			Epoxy resin adhesive  & 2100 \\ 
			\hline
			Air (cracks and debonds)  & 340 \\ 
			\hline
			Kissing bond  & 1500 \\ 			
			\hline
			\hline
	\end{tabular}}
\end{table}
\subsubsection{Modelling procedures}\label{sec:modelling procedures}

In this work, the implicit time-domain staggered-grid finite difference (FD) scheme using second-order in time and eighth-order in space is used to solve the acoustic wave equation (see Eq.~(\ref{eq:forward_modelling})). Each velocity model in this work consists of a uniform grid with a spacing of 0.1 mm.~This guarantees the calculation accuracy which requires at least four grid points per shortest wavelength~\cite{Hustedt2004}. The space domain is surrounded by perfectly matched layers to avoid reﬂections coming from the left, right and bottom edges. The input Ricker signal with a central frequency of 5 MHz is excited and monitored by a linear phased array with 64 equally spaced elements placed on the top surface of the model, as shown in Figs.~\ref{fig:configuration} (a) and (b). In Type I and Type II models, the pitch of elements in the array is 0.6 mm and 0.8 mm, respectively. To simulate the working process of the ultrasonic phased array imaging, the 64 transmitter elements are fired successively while all the receiver elements are working simultaneously. Then all transmitter-receiver combinations in the FMC data are produced. Note that only the ultrasonic L-wave is excited and recorded, and shear-waves (S-waves) and mode conversions between L-waves and S-waves are not considered in this paper.

\subsubsection{Testing dataset}\label{sec:Testing data}

In the testing dataset, the ultrasonic velocity models of multi-layered bonded composites have similar distributed structures as the examples in the training dataset, because the proposed method in this paper is the supervised learning method. However, the test examples are not included in the training dataset and are unknown in the predicting process. The ultrasonic FMC data as inputs in the predicting process are obtained from simulations. In this work, 20 examples of Type I and 20 examples of Type II are used to evaluate the proposed method. 

\subsection{Inversion results}\label{sec:specimen 1}
\subsubsection{Type I}\label{sec:Type I}

\begin{figure*}[!htbp]
	\center
		\includegraphics[width=0.8\textwidth]{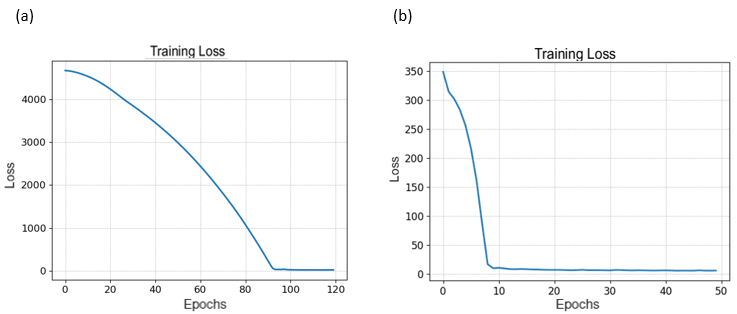}
		\caption{Training and loss history (Loss outlined in Algorithm \ref{alg1} vs Epoch) of (a) Type I and (b) Type II. \label{fig:trainloss1}}
\end{figure*}

In this section, the inversion is performed for Type I with debond and kissing bond defects in the 1 mm-thick adhesive layer as well as cracks in the top layer made of titanium, as shown in Fig.~\ref{fig:configuration} (a). In the training process, the learning rate of the Adam is set as~$10^{-3}$ for the velocity models with grid points of 401 $\times$ 201, and the number of epochs is chosen as 120 according to Fig.~\ref{fig:trainloss1} (a). Note that the loss in Fig.~\ref{fig:trainloss1} is computed pixel-based (see Algorithm \ref{alg1}) and, namely, it is the sum of all pixel differences. The weight initialization strategy is described in~\cite{He2015} and the initial value of the bias is set to zero. Other parameters used in the training process are illustrated in Table \ref{tab:training_parameters_Type}. After training, 20 examples from the testing dataset are used to test the performance of the FCN-based ultrasonic inversion method. 

\begin{table}[!t]
	\caption{Parameters used in the training process of Type I and Type II}
	\label{tab:training_parameters_Type}
	\centering
	{\begin{tabular}{lcccccc}
			\hline
			\hline
			& Learning rate   & Epoch  & Batch size  & Number of \\
			& Adam algorithm ($\alpha$) & (E) & (h) & training data (N)
			\\
			\hline
			Type I  & $10^{-3}$ & 120 & 10 & 800 \\
			\hline
		    Type II  & $10^{-2}$ & 50 &  10 & 800 \\ 
			\hline
			\hline
	\end{tabular}}
\end{table}

\begin{figure*}[!htbp]
	\center
		\includegraphics[width=0.9\textwidth]{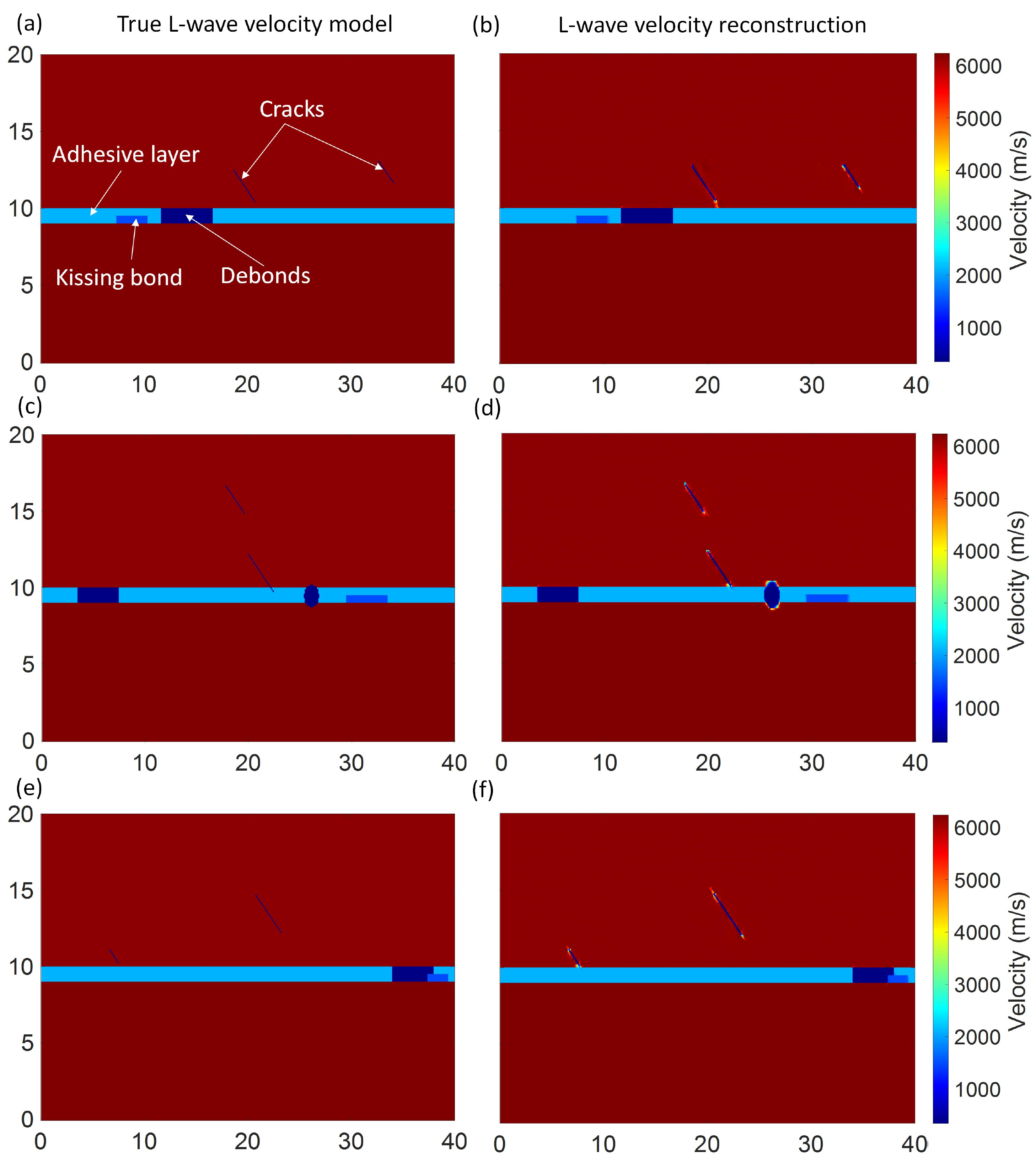}
		\caption{Reconstructions of Type I models. Cracks are contained in the top layer of titanium and the debond and kissing bond defects are hidden in the adhesive layer, as shown in the true L-wave velocity models of (a). (b) shows the FCN-based ultrasonic inversion reconstruction. The unit in this figure is mm. \label{fig:TypeI_predicted_models}}
\end{figure*}

Three representative true L-wave velocity models of Type I from the testing dataset are shown in Figs.~\ref{fig:TypeI_predicted_models} (a), (c) and (e), while Figs.~\ref{fig:TypeI_predicted_models} (b), (d) and (f) give the corresponding velocity reconstructions using the FCN-based ultrasonic inversion method. It can be seen that high-quality quantitative images of large velocity contrast are achieved, and the locations and the shapes of the defects are reconstructed well. The length of cracks and the size of the debond with a circular shape (Fig.~\ref{fig:TypeI_predicted_models} (d)) are slightly larger than the true models. A possible reason could be that the proposed method is based on the acoustic wave equation, meaning that the S-waves and mode conversions are not considered in this work. Although the elastic wave equations are more suitable to describe the material behaviour than the acoustic wave equations, the acoustic equations can yield good kinematic L-wave approximations to the elastic wave equations~\cite{Zhang2005}. This not only reduces the computational costs, but is also convenient for inverse algorithms. 

\subsubsection{Type II}\label{sec:Type II}

Type II models with multiple debonds in the 0.1 mm-thick adhesive layer compared to the 1 mm-thick adhesive layer, as shown in Fig.~\ref{fig:configuration} (b), are used to further validate the reconstruction capability of the FCN-based ultrasonic inversion method. The learning rate of the Adam is $10^{-2}$ for all velocity models with grid points of 701 $\times$ 201, and the number of epochs is 50 according to Fig.~\ref{fig:trainloss1} (b). More details on the parameters used in the training process are shown in Table \ref{tab:training_parameters_Type}.

\begin{figure*}[!htbp]
	\center
		\includegraphics[width=0.9\textwidth]{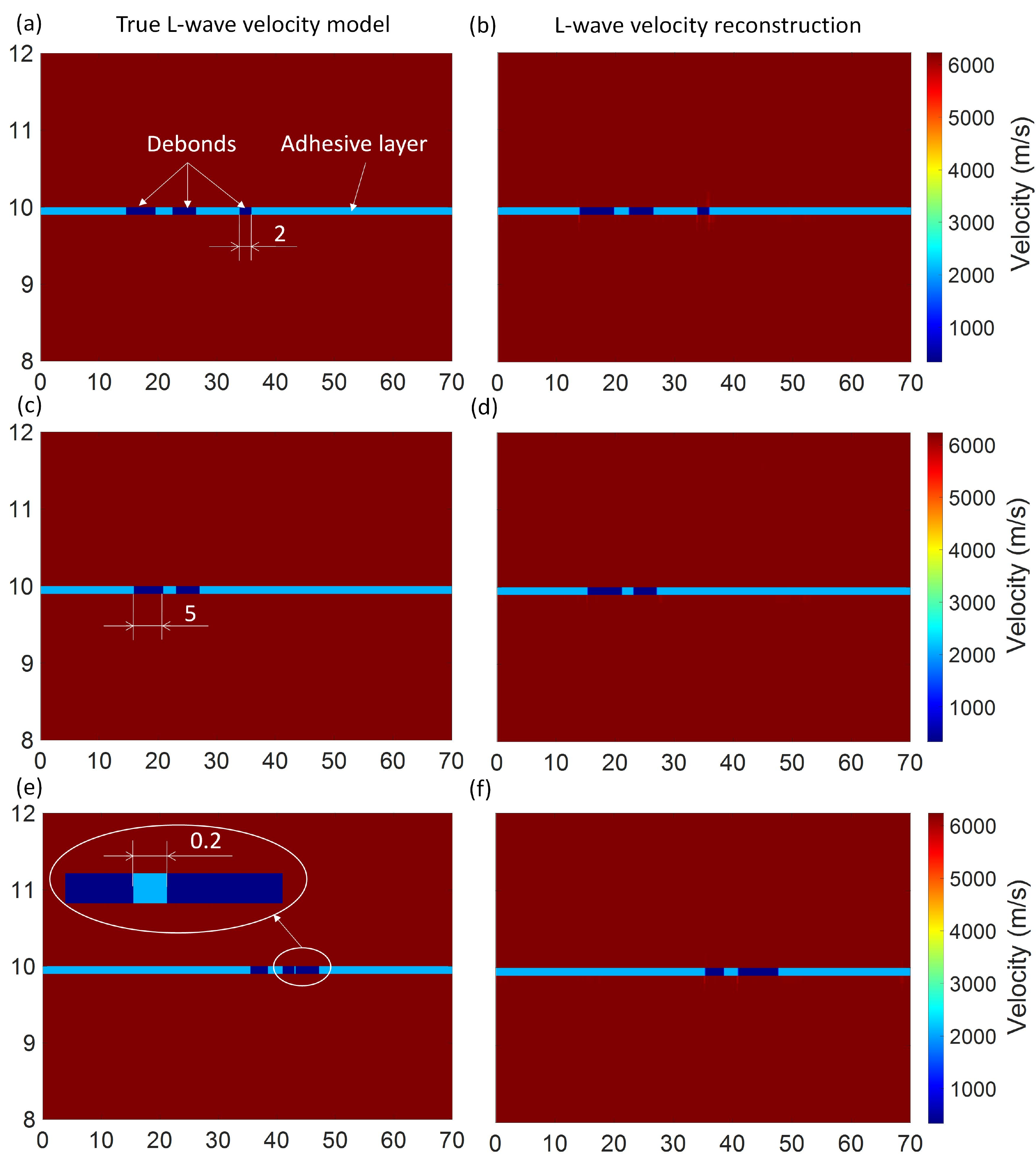}
		\caption{Reconstructions of Type II models. Multiple debond defects are hidden in the adhesive layer, as shown in the true L-wave velocity models of (a). (b) gives the FCN-based ultrasonic inversion reconstruction. The unit in this figure is mm. \label{fig:TypeII_predicted_models}}
\end{figure*}

Figs.~\ref{fig:TypeII_predicted_models} (a), (c) and (e) show three representative examples of Type II with different debond lengths and spacing between debond defects from the testing dataset. The lengths of the debonds range from 2 mm to 5 mm, and the minimum spacing between two adjacent debonds shown in Fig.~\ref{fig:TypeII_predicted_models} (e) is 0.2 mm. The reconstruction results of multiple debonds are shown in Figs.~\ref{fig:TypeII_predicted_models} (b), (d) and (f). Figs.~\ref{fig:TypeII_predicted_models} (b) and (d) clearly show that the FCN-based ultrasonic inversion reconstructions of the high contrast debonds are accurate, producing a good estimate of different lengths and locations. Fig.~\ref{fig:TypeII_predicted_models} (f) presents the reconstruction of three debonds with the minimum spacing of 0.2 mm. This reconstruction provides a good overall estimate of the debonds, and it is quite close to the true values (Fig.~\ref{fig:TypeII_predicted_models} (e)). However, the reconstruction fails to capture the adjacent debonds with a sufficiently small spacing of 0.2 mm. Besides the possible reason mentioned above, the relatively small example size of the training dataset (800), which may not contain the similar examples with very small spacing between two adjacent debonds, has a great impact on the predicted results. 

In general, the reconstruction results of Type I and Type II shown in this work confirm that the proposed method has the ability to reconstruct the velocity model of adhesively bonded composites containing high contrast defects directly from the measured ultrasonic FMC data. 

\section{Discussion}\label{sec:Discussion}

The numerical results presented in this paper for defect detection have shown that the FCN-based ultrasonic inversion method can be a useful tool to achieve an accurate quantitative reconstruction of multi-layered bonded composites. Compared with other conventional inversion methods, the reconstruction costs of this method are negligible once the trained network is obtained in the training process. Additionally, this method neither requires an initial velocity model, nor does it exhibit any cycle-skipping problems~\cite{Virieux2009}. However, there are several factors that can affect the performance of the proposed method, such as the selection of training datasets, the parameters (learning rate, batch size, epoch, et al.) used in the training process as well as the architecture of the network. For example, the proposed method is based on the supervised-learning network, and the capability of this network relies on the training dataset. In addition, the velocity models, which can be accurately predicted in the testing dataset, should exhibit similarly distributed structures as the velocity models used in the training dataset. Generally, with a larger amount of the training datasets, it can result in more accurate trained network~\cite{Bengio2012}, which can help to achieve a more accurate velocity reconstruction. However, this in turn also increases the computational effort to train the network. The influence of the training dataset on the proposed method will be further investigated along with the physical experimental dataset used in the training process. 

\section{Conclusion}\label{sec:Conclusion}

In this paper, the fully convolutional network(FCN)-based ultrasonic inversion method is proposed for quantitative reconstruction of high contrast defects hidden in multi-layered bonded composites. This supervised end-to-end method bypasses the need to carefully select the appropriate initial velocity model and inversion parameters. It uses a network to transform the raw-input ultrasonic data directly to the velocity reconstruction. The trained network is applied to reconstruct velocity models of multi-layered bonded composites from the testing dataset. In this work, the effectiveness of the proposed technique is tested using two types of composites with disbond and kissing bond defects in 1 mm-thick adhesive layer and cracks in the titanium layer, as well as with multiple debonds in the 0.1 mm-thick adhesive layer. The numerical results show a good consistency with the true values. 

From this, we conclude that the FCN‐based inversion method has an attractive potential for reconstructing high contrast defects. The validation in this paper is based on simulations. This is intended as a first step towards the application of the FCN-based ultrasonic inversion method in physical experiments and field tests.

\bibliographystyle{IEEEtran}
\bibliography{IEEEabrv,mybibfile}

\end{document}